\documentclass[12pt]{article}
\usepackage{amscd}
\usepackage{verbatim}
\usepackage{amssymb}
\usepackage{amsmath}
\begin{document}
\begin{center}
\vspace{24pt} { \large \bf A proposal for studying off-shell
stability of vacuum geometries in string theory}  \\ \vspace{30pt}
{\bf V Suneeta} \footnote{vardarajan@math.ualberta.ca}\\
\vspace{24pt} %%
{\footnotesize Dept of Mathematical and Statistical Sciences \\ and
\\ The Applied Mathematics Institute,\\ University of Alberta,
Edmonton, AB, Canada T6G 2G1.}
\end{center}
\date{\today}
\bigskip

\begin{center}
{\bf Abstract}
\end{center}
\noindent We briefly review studies of off-shell stability of vacuum
geometries in semiclassical gravity. We propose a study of off-shell
stability of vacua in string theory by a distinct, though somewhat
related approach --- by studying their stability under suitable
world-sheet sigma model renormalization group (RG) flows. Stability
under RG flow is a mathematically well-posed and tractable problem
in many cases, as we illustrate through examples. The advantage is
that we can make definite predictions about late time behaviour and
endpoints of off-shell processes in string theory. This is a
contribution to the proceedings of Theory Canada 4, CRM Montreal.
\newpage

\section{Introduction} \setcounter{equation}{0} Studies
of off-shell stability in the context of semiclassical gravity are
not new. It has been observed that off-shell perturbations of
classical solutions that make the spacetime action negative lead to
instabilities (in the semiclassical approximation) in quantum
gravity \cite{page}, \cite{GPY}. These instabilities are expected to
cause quantum tunnelling from one classical configuration to another
through off-shell configurations. An example of these is given by
the Euclidean Schwarzschild instanton. There are perturbations of
this background that make the spacetime action negative. One such
mode, which is non-normalizable is expected to cause tunnelling
between the large Schwarzschild black hole and hot flat space.

A similar exercise can be done for gravity in a cavity which is kept
at fixed temperature in contact with a heat bath \cite{WY},
\cite{BA}, \cite{TP}. In this case, the action has three saddle
points, hot flat space, the small (Euclidean) Schwarzschild black
hole and the large black hole. As has been discussed in \cite{HW},
the small black hole has an off-shell perturbation that makes the
action negative, and is therefore unstable. It is expected to
mediate in a quantum tunnelling between the large black hole and hot
flat space. Furthermore, what is interesting is the link between
local thermodynamic instability of the Lorentzian black hole (e.g.,
the small Schwarzschild black hole) and the off-shell perturbative
instability of its Euclidean counterpart \cite{WY}, \cite{reall}. As
Reall points out, this perturbative instability is related to a
classical Gregory-Laflamme instability of the black brane whose
dimensional reduction (and Wick rotation) gives the Euclidean
Schwarzschild black hole. The generalization of this analysis to
dimensionally reduced black branes in supergravity also follows by
looking at a suitable action (possibly involving other fields as
well) and seeing if there are off-shell perturbations of a classical
solution that can make the action negative \cite{reall}.

In this article, we discuss a natural notion of `off-shell
stability' in closed string theory that is somewhat related to but
distinct from the above analysis. One would expect to study
off-shell stability of a geometry in string theory by considering
off-shell configurations in an appropriate string field theory (for
a recent review of string field theory, see \cite{rastelli}).
However, at present, the study of off-shell stability in closed
string field theory is a very hard problem. We will therefore work
in the low energy approximation, and use the world-sheet sigma model
approach. Then we can view an on-shell geometry (a string vacuum) as
arising as a fixed point of the renormalization group (RG) flow of
the sigma model. As is well known, the sigma model corresponding to
this geometry is then a conformal field theory (CFT). In many
examples, it is easy to consider generic relevant perturbations of
this CFT that are also tachyonic. Relevant perturbations induce RG
flows of the sigma model. A perturbation that is tachyonic is
indicative of a genuine off-shell instability in string field
theory, and it is expected that there will be tachyon condensation
(an off-shell process). Tachyon condensation generically leads to a
change in geometry (for a review, see \cite{HMT}). Thus it is
natural to ask whether the two processes induced by the relevant
tachyonic perturbation are related; i.e., whether the off-shell
geometries that solve the RG flow equations are the same as those
involved in the geometry change due to tachyon condensation
\cite{HMT}. There is some evidence that RG flow approximates the
geometry change due to tachyon condensation, with the endpoints
being the same in many cases. An example is the orbifold flow $C/Z_n
\rightarrow C $ expected from tachyon condensation \cite{APS},
\cite{GHMS}, \cite{OZ} --- there is an exact solution to RG flow
with precisely the same endpoints \cite{Cao}, \cite{GHMS}. In open
string theory, there is an even more striking example --- the Sen
conjecture for unstable D-branes has been verified both using string
field theory and world-sheet RG flows (see \cite{HMT} for a review).
In fact, the idea of exploring the configuration space in string
(field) theory by studying RG flows of world-sheet sigma models is
very old and was proposed in the late 1980s by Banks, Martinec, Vafa
and others \cite{BM}, \cite{Vafa}. It is hoped that future progress
in string field theory will lead to a rigorous result relating
evolution of off-shell perturbations in the theory to RG flow
solutions in a suitable approximation.

Motivated by the above discussion, we propose to investigate
off-shell stability of closed string vacua by studying their
stability under suitable RG flows. (In)stability under RG flow is
expected to be indicative of (in)stability under off-shell
perturbations in string field theory (at least for a class of
off-shell perturbations, in light of the above evidence). The
advantage of the RG flow approach is that in many cases, the
question of stability under RG flow is a mathematically well-posed
problem. In this article, we first introduce some simple RG flows of
string theory. We then discuss stability under suitable RG flows for
different geometries of interest in string theory, highlighting the
stability techniques used. Finally we comment on the possible
applications of these stability results. \section{ RG flows in
closed string theory, linear and geometric stability under RG flow }
\setcounter{equation}{0} The $\beta$ functions of the closed string
world-sheet sigma models are obtained as a perturbative expansion in
powers of the square of the string length $\alpha'$. Truncating this
expansion to first order in $\alpha'$ is a valid approximation as
long as the magnitude of curvature of the target space is much less
than $1/\alpha'$. The simplest first-order RG flow is the {\em Ricci
flow}, which is well-studied in mathematics and has been used
successfully to prove the Poincar\'e conjecture. This is a flow for
the metric of the target space, $g_{ij}$, with respect to the RG
flow parameter $t$, and what is physically significant is the flow
of geometries (i.e., of metrics mod diffeomorphisms). Ricci flow is
not form-invariant under arbitrary $t$ dependent diffeomorphisms, so
all flows related to each other by diffeomorphisms generated by a
vector field $V$ are physically equivalent. We write the most
general flow in this class, also called the {\em Ricci-de Turck
flow} as
\begin{eqnarray}
\frac{\partial g_{ij}}{\partial t} = -\alpha' (R_{ij} + \nabla_i V_j
+ \nabla_j V_i). \label{2.1} \end{eqnarray}

This is the RG flow for the metric when all other fields are set to
zero. In the case when there is a non-zero dilaton $\Phi$, we can
write $V_i = \nabla_i \Phi$ in (\ref{2.1}) to get the flow for the
metric. The flow for the dilaton is then
\begin{eqnarray}
\frac{\partial \Phi}{\partial t} = \frac{\alpha'}{2} ( \Delta \Phi -
2 |\nabla \Phi |^2 ). \label{2.2} \end{eqnarray}

In the presence of a nonzero $B$ field, we have a more complicated
set of coupled equations which can be found, for e.g., in
\cite{Polchinski}.

Fixed points of the Ricci-de Turck flow are candidates for being
string vacua. It is well-known that when the manifold is compact,
the only fixed points of Ricci-de Turck flow are Ricci flat
\cite{bour} (this need not be true for RG flows involving other
fields as well, see, for instance \cite{gs}). On noncompact
manifolds, non-Ricci flat fixed points can exist, an example being
the Witten black hole \cite{MSW},\cite{witten}. Given such a fixed
point of (\ref{2.2}), the question of stability of the fixed point
under Ricci flow for small perturbations is mathematically
well-posed and can be investigated.

On compact manifolds, the technique used to investigate stability by
Cao, Hamilton and Ilmanen \cite{CHI} applies some results derived by
Perelman as part of his proof of the Poincar\'e conjecture
\cite{perelman}. Briefly, there is a diffeomorphism-invariant
functional (the $\lambda$ functional) that is constant only at fixed
points of geometry under Ricci flow and monotonically increasing
otherwise. Now, let us consider a fixed point with metric $g$ (for
which we can evaluate this functional, we call this $\lambda(g)$)
and perturb this fixed point at some initial value of RG parameter
$t_0$. Denote the perturbed metric at $t_0$ by $g^{p}(t_0)$.
Evaluate $\lambda$ for $g^{p}(t_0)$, denoted by $\lambda
(g^{p}(t_0))$. The sign of $\lambda (g^{p}(t_0)) - \lambda(g)$ can
be used to check for stability of the fixed point
--- if this sign is positive, then since $\lambda$ for the perturbed
geometry monotonically increases, the perturbed geometry will not
approach the fixed point; and if this sign is negative, it will.
$\lambda (g^{p}(t_0)) - \lambda(g)$ is then evaluated in a
linearized approximation assuming that $g^{p}(t_0)$ is a small
perturbation of $g$ (note that this is only required at initial
$t=t_0$, not all along the flow). For Ricci-flat metrics, this sign
is related to the sign of the spectrum of the Lichnerowicz laplacian
on the manifold. It has been shown that with the exception of
scaling and other flat directions associated to the moduli space of
the $n$-torus, the $n$-torus is linearly stable under Ricci flow.
This has also been proven by other methods (maximal regularity) in
\cite{GIK}. This can be generalized to RG flow with a B-field
\cite{osw1}. Similarly, the K\"ahler-Ricci-flat metric on K3 is
linearly stable except for a centre manifold of flat directions (and
scaling) \cite{GIK}, \cite{sesum}.

Another notion of stability discussed in \cite{CHI} is that of {\em
geometric stability}. Consider a geometry that is a solution to RG
flow but not a fixed point. One can still analyze whether this
solution is an {\em attractor} for nearby metrics (up to
diffeomorphisms and scaling). From the point of view of sigma model
quantum field theory, such special geometries are interesting. But
what would be the use of such a stability result in string theory?
At the very least, we can examine the sigma model corresponding to
the limiting geometry as $t \rightarrow \infty$ of this special
solution since it is the natural end-point of many nearby RG flows.
Thus, given the conjecture that RG flows approximate off-shell
processes in string theory, this end-point may give insights into
the end-points of off-shell processes. The quantum field theory at
the end-point need not be a conformal field theory
--- an example when it is not is that of $S^n$. In \cite{CHI}, it is shown
that $S^n$ is geometrically stable. Therefore, we can ask what the
sigma model corresponding to $S^n$ flows to as $t \rightarrow
\infty$. It is well-known that $S^n$ shrinks under Ricci flow
\cite{chow}, and when its curvature is of order $1/\alpha'$, sigma
model perturbation theory is no longer reliable. For the case of
$S^2$, it is known (both using techniques from integrable models and
numerical simulations) what the limiting world-sheet quantum field
theory is
--- it is a free field theory of {\em massive} fields \cite{FOZ},
\cite{Friedan}.

There is also the possibility that an attractor solution to an RG
flow may be a genuine fixed point of another RG flow in string
theory, possibly with the addition of other fields. For example,
hyperbolic $3$-space $H^3$ (or the BTZ black hole, which is a
quotient geometry of $H^3$) is not a fixed point of the flow
(\ref{2.1}), but is a fixed point of RG flow with a non-zero $B$
field \cite{hw}, \cite{kaloper}. In such a case, a geometric
stability result under the flow (\ref{2.1}) is expected to be
indicative of a linear stability result under RG flow with a $B$
field, at least with respect to metric perturbations. Therefore, the
question of whether $H^3$ is geometrically stable under Ricci flow
is worth investigating. In fact, we know that $H^n$ (i.e., Euclidean
anti-de Sitter space $AdS_n$) arises in many contexts in string
theory. There are supergravity solutions of the form $AdS_p \times
S^q$ for specific integers $p, q > 0$. There are world-sheet
descriptions as well for these geometries, and it is expected that
they are fixed points of some RG flow that includes the effects of
the RR fields. However the $\beta$ function for the world-sheet
theory is not known. So if we want to investigate stability of, say,
$AdS_p \times S^q$ (or the Wick-rotated geometry $H^p \times S^q$),
then the best we can do is investigate its geometric stability under
Ricci flow. As discussed, it is known that $S^q$ is geometrically
stable. So a result on geometric stability of $H^n$ under Ricci flow
would at least imply stability of $H^p \times S^q$ under a special
class of metric perturbations and would be the first step towards a
complete stability analysis of $H^p \times S^q$. We also wish to
mention that the classical stability of $AdS^p \times S^q$ in
supergravity has been well-studied \cite{freed}.

In the next section, we review a result on geometric stability of
$H^n$ under Ricci flow that is derived in \cite{vs}. In general,
investigation of linear or geometric stability for non-compact
manifolds is a difficult problem. This is because the analogue of
the $\lambda$ functional for noncompact manifolds only exists in
certain cases (with some conditions on sign of curvature
\cite{osw}). So techniques such as that of Cao, Hamilton and Ilmanen
cannot be used. We briefly mention some noncompact geometries of
interest and known stability results for these geometries. We then
highlight a new stability technique motivated by relativity
\cite{am} and review how it is used to analyze stability of $H^n$ in
\cite{vs}.
\section{Stability of noncompact geometries}
 \setcounter{equation}{0}We first discuss linear stability results
for noncompact fixed points of (\ref{2.1}). The most obvious example
is flat space. In this case, with conditions on the curvature and
its decay asymptotically, stability results are derived in
\cite{shi}. A more recent result is found in \cite{SSS}, where the
metric at initial $t$ is required to be a perturbation of the flat
metric on $R^n$, and as well is required to satisfy some asymptotic
conditions.

There exist nontrivial fixed points to (\ref{2.1}) with $V_j \neq 0$
\cite{Cao1}, \cite{bryant}. Such solutions are referred to as {\em
steady Ricci solitons} in Ricci flow literature. Stability results
for some steady solitons can be found in \cite{cs} (with suitable
asymptotic decay conditions imposed on the perturbation). One
example of a Ricci soliton well-studied in physics is the
(Wick-rotated) Witten black hole \cite{MSW}, \cite{witten}, given by
the metric
\begin{eqnarray}
ds^2 = dr^2 + \tanh^2 r d\theta^2, \label{3.1}
\end{eqnarray}
with $V_i = \nabla_i \Phi$ and $\Phi = \log(\cosh r)$. The Witten
black hole arises as the backward limit (in a specific sense) of a
collapsing solution to first-order RG (i.e., Ricci) flow, the
sausage model \cite{FOZ} (in mathematics literature, this is called
the Rosenau solution, see \cite{chow} for details). This may
indicate the instability of the Witten black hole under specific
perturbations. Thus the stability of the Witten black hole remains
an open question.

Finally, we discuss the geometric stability of $H^n$ (Euclidean
$AdS_n$) under Ricci flow \cite{vs}. We use this example to
introduce a new technique motivated by recent work of Andersson and
Moncrief \cite{am} in relativity. We first outline the technique.
Consider a fixed point of an RG flow --- a manifold with metric
$g_{ij}$. Now, consider a perturbation of this fixed point, with the
perturbed geometry having metric $g_{ij}^{p} = g_{ij} + h_{ij}$,
where $h_{ij}$ is assumed small. Then we can work with the
linearized RG flow for the perturbation $h_{ij}$. We now define the
{\em Sobolev norm} of the perturbation $\parallel h
\parallel_{k,2}$,
\begin{eqnarray}
\left ( \parallel h \parallel_{k,2} \right )^2 &=& \int_M |h_{ij}|^2
~dV + \int_M |\nabla_{p_1} h_{ij}|^2 ~dV + ....\nonumber \\&&...+
\int_M |\nabla_{p_1}....\nabla_{p_k} h_{ij} |^2 ~dV.\label{3.2}
\end{eqnarray}
In the above expression, $|h_{ij}|^2$, for example, is
$h^{ij}h_{ij}$, and indices are raised with respect to the
background metric $g$. We assume that this norm is defined for our
perturbation at initial $t$. Then, if we can prove that the Sobolev
norm of the perturbation is decaying under the linearized flow,
i.e., $\parallel h
\parallel_{k,2} \rightarrow 0$ as $t \rightarrow \infty$, then in
many cases, we can prove using certain {\em Sobolev inequalities}
that the perturbation and its derivatives decay {\em pointwise}
everywhere on the manifold. We would like to illustrate this by
studying geometric stability of $H^n$ under Ricci flow. Recall that
$H^n$ is not a fixed point of Ricci flow --- it expands uniformly
under the flow. We are interested in whether its perturbations decay
and if the perturbed geometry approaches the expanding $H^n$
geometry under Ricci flow. However, the technique we just outlined
was for studying stability of a {\em fixed point}. This does not
pose a problem, since solutions to the Ricci flow
\begin{eqnarray}
\frac{\partial \tilde g_{ij}}{\partial \tilde t} = - \alpha' \tilde
R_{ij} \label{3.3}
\end{eqnarray}
are related to those of the flow
\begin{eqnarray}
\frac{\partial g_{ij}}{\partial t} = - \alpha' [ R_{ij} + c g_{ij} ]
\label{3.4}
\end{eqnarray}
by the rescalings
\begin{eqnarray}
\tilde t = \frac{1}{\alpha' c}~~ e^{\alpha' c t}, \nonumber \\
\tilde g_{ij} = e^{\alpha' c t} g_{ij}. \label{3.5}
\end{eqnarray}
Writing the Riemann tensor of $H^n$ as $R_{ijkl} = - \frac{1}{a}
(g_{ik}g_{jl} - g_{il}g_{jk})$ where $a > 0$ is a constant, $R_{ij}
= - \frac{(n-1)}{a} g_{ij}$ and $R = - \frac{n(n-1)}{a}$ ($a$
characterizes the magnitude of curvature). We then see that $H^n$ is
a fixed point of the flow (\ref{3.4}) with $c = \frac{(n-1)}{a}$. It
is easy to see from (\ref{3.5}) that linear stability of $H^n$ under
the flow (\ref{3.4}) implies its geometric stability under the Ricci
flow (\ref{3.3}). To prove linear stability under (\ref{3.4}), we
first consider the linearized flow of the perturbation $h_{ij}$ of
$H^n$, which (after choosing an appropriate gauge), we write as
\begin{eqnarray}
\frac{\partial h_{ij}}{\partial t} = (\Lambda h)_{ij} \label{3.6}
\end{eqnarray} where
$(\Lambda~ h)_{ij} = \frac{\alpha'}{2} \left [ (\Delta_{L} h)_{ij} -
2 c h_{ij} \right ]$  and $\Delta_{L}$ is the Lichnerowicz Laplacian
operator that also appeared in studies of off-shell stability in
semiclassical gravity \footnote{ We use the sign convention in
mathematics literature for $\Delta_{L}$. This differs by a sign from
the convention used by physicists.}. We need to consider only
transverse perturbations as the divergence part of a perturbation is
`pure gauge', similar to perturbation analysis in classical
relativity. We then define the following `energy' integrals
\begin{eqnarray}
E^{(K)} = \int_M \left |(\Lambda^{(K)} h)_{ij} \right |^2 ~ dV ,
\label{3.7}
\end{eqnarray}
where the notation $( \Lambda^{(2)} h )_{ij} = (\Lambda \Lambda
h)_{ij}$, for example (denote $(\Lambda^{(0)}h)_{ij} = h_{ij}$).
With appropriate asymptotic conditions on the perturbation, we can
then derive the following bound for $n > 2$:
\begin{eqnarray}
 \frac{dE^{(K)}}{dt} \leq - \alpha' \frac{(n-2)}{a} E^{(K)} \label{3.8}
\end{eqnarray}

Then, we can show that this bound implies that (with the appropriate
asymptotic conditions) the Sobolev norm of the perturbation decays
as $t \rightarrow \infty$ (see \cite{vs} for the details). For
$n=2$, we only need to consider the flow of the trace of the
perturbation. We can prove similarly that the Sobolev norm of the
trace decays under the flow. Choosing the Poincar\'e unit ball
metric for $H^n$, we have the following Sobolev inequality derived
by Andersson \cite{andersson}, for $s
> n/2$ ($s$ is a nonnegative integer):
\begin{eqnarray}
\parallel h \parallel_{C^k} ~ \leq C \parallel h \parallel_{k+s,~ 2} .
\label{3.9}
\end{eqnarray}
$\parallel h \parallel_{C^k}$ refers to the $C^k $ norm of the
perturbation. Let $\alpha = (\alpha_{1},....\alpha_{j})$ be the
$j$-tuple of non-negative integers $\alpha_{i}$. We call $\alpha$ a
multi-index. $\nabla^{\alpha} =
\nabla_{1}^{\alpha_{1}}...\nabla_{j}^{\alpha_{j}}$ is a differential
operator of order $|\alpha| = \sum_{i=1}^{j} \alpha_{i}$. The $C^k$
norm of u is defined as
\begin{eqnarray}
\parallel u \parallel_{C^k} = \max_{0 \leq |\alpha| \leq k}~\sup_{x \in M}
|\nabla^{\alpha} h(x)|, \label{3.10} \end{eqnarray} where, for
instance, $|h(x)| = [h^{ij}(x)h_{ij}(x)]^\frac{1}{2}$ is the usual
pointwise tensor norm defined with respect to the background
hyperbolic metric. Then the fact that the Sobolev norm of the
perturbation decays as $t \rightarrow \infty$ implies from
(\ref{3.9}) that the perturbation and its derivatives decay
pointwise on the manifold. This proves the geometric stability of
$H^n$ under Ricci flow.

This technique can be easily generalized to other RG flows. $H^3$ is
the fixed point of RG flow with a $B$-field. Thus a natural
extension would be to study the stability of $H^3$ under this RG
flow. Another question is the stability of quotient geometries of
$H^n$. The most interesting of these is the BTZ black hole which is
obtained as a quotient of $H^3$. A stability result for the BTZ
black hole can probably be derived, but the study of boundary terms
and asymptotic conditions will differ from our example. At this
stage, we are unable to study general perturbations of $H^p \times
S^q$, since this geometry evolves under Ricci flow (the $H^p$ part
expands and $S^q$ shrinks), and we cannot find a flow with this
geometry as fixed point, such that its solutions are related to
those of Ricci flow by rescalings, similar to (\ref{3.4}). However,
these examples show that stability analysis of geometries under RG
flows is in many cases, a well-posed problem and offers a tractable
method of studying off-shell stability in string theory.

Finally, we observe that both in the off-shell stability studies in
semiclassical gravity and in stability studies under the simplest RG
flows, the Lichnerowicz operator (particularly the sign of its
spectrum) plays a crucial role. We could study the stability of the
same Euclidean black hole and black brane geometries considered by
Reall under suitable RG flows. It is likely that a geometry that is
semiclassically unstable would also be unstable under RG flow since
both notions of off-shell stability rely on the spectrum of the
Lichnerowicz operator. The advantage of the RG flow approach is that
we can follow the evolution of the perturbation to late RG times,
and make a prediction about the endpoint geometry. In cases where a
geometry is linearly stable, it may be possible to extend this to a
notion of {\em nonlinear} stability under RG flow, analogous to
recent nonlinear stability results (for compact hyperbolic space) in
relativity \cite{am}. Thus we hope that this program will offer
insights into late time behaviour and endpoints of off-shell
processes in string theory.
\section{Acknowledgements}
This is a contribution to the Theory Canada 4 Proceedings, and is a
review of my recent work \cite{vs}. I would like to thank the
organizers for a very enjoyable and stimulating conference.

\end{document}